\documentclass[12pt, a4paper]{article}

\usepackage{amsmath}
\usepackage{amsfonts}
\usepackage{amssymb}
\usepackage{bbm}
\usepackage{verbatim}
\usepackage{dsfont}
\usepackage{booktabs}
\usepackage{slashed}

\usepackage{epsfig}
\usepackage{color}
\usepackage[table]{xcolor}
\usepackage{graphicx}
\usepackage[footnotesize]{caption}
\usepackage[listofformat=empty,subrefformat=empty]{subfig}

\usepackage{float}
\usepackage{overpic}

\usepackage{enumerate}
\usepackage{hhline}
\usepackage{multirow}

\usepackage{cite}
\usepackage{xspace}
\usepackage{setspace}

\usepackage[pdftitle={Split symmetries},
  pdfauthor={Wilfried Buchmuller, Markus Dierigl, Fabian Ruehle, Julian Schweizer},
   pdfsubject={orbifolds, Wilson lines, flux, anomalies, GUT},
   bookmarksopen, bookmarksnumbered, bookmarksopenlevel=2, colorlinks=false, linkcolor=blue, citecolor=blue, urlcolor=blue]{hyperref}

\renewcommand{\tfrac}{\genfrac{}{}{}1}

\begin{document}

\thispagestyle{empty}

\begin{flushright}
DESY-15-095
\end{flushright}
\vskip .8 cm
\begin{center}
{\Large {\bf Split symmetries}}\\[10pt]

\bigskip
\bigskip 
{
{\bf{Wilfried Buchmuller}\footnote{E-mail: wilfried.buchmueller@desy.de}},
{\bf{Markus Dierigl}\footnote{E-mail: markus.dierigl@desy.de}},  
{\bf{Fabian Ruehle}\footnote{E-mail: fabian.ruehle@desy.de}},
{\bf{Julian~Schweizer}\footnote{E-mail: julian.schweizer@desy.de}}
\bigskip}\\[0pt]
\vspace{0.23cm}
{\it Deutsches Elektronen-Synchrotron DESY, 22607 Hamburg, Germany }\\[20pt] 
\bigskip
\end{center}

\begin{abstract}
We consider six-dimensional supergravity with gauge group $SO(10)\times U(1)_A$, compactified on the orbifold $T^2 / \mathbb{Z}_2$. Three quark-lepton generations arise as zero modes of a bulk $\bf16$-plet due to magnetic flux of the anomalous $U(1)_A$. Boundary conditions at the four fixed points break $SO(10)$ to subgroups whose intersection is the Standard Model gauge group. The gauge and Higgs sector consist of ``split'' $SO(10)$ multiplets. As consequence of the $U(1)_A$ flux, squarks and sleptons are much heavier than gauge bosons, Higgs bosons, gauginos and higgsinos. We thus obtain a picture similar to ``split supersymmetry''. The flavor structure of the quark and lepton mass matrices is determined by the symmetry breaking at the orbifold fixed points.
\end{abstract}

\newpage 
\setcounter{page}{2}
\setcounter{footnote}{0}

{\renewcommand{\baselinestretch}{1.3}
\onehalfspacing

\section{Introduction}
\label{sec: Introduction}

Fermions and bosons play a very different role in the Standard
Model. It is remarkable that quarks and leptons form three copies of complete
multiplets of a grand unified (GUT) group, $SU(5)$ or $SO(10)$, whereas
gauge and Higgs bosons are single, incomplete, ``split'' multiplets.
In the following we shall propose a model where this difference is
explained by connecting GUT symmetry breaking and supersymmetry
breaking: Scalar quarks and leptons are very heavy because they belong
to complete GUT multiplets, whereas supersymmetry breaking is small for
gauge and Higgs fields since they form incomplete GUT multiplets.
One is thus led to a picture similar to ``split supersymmetry'' \cite{ArkaniHamed:2004fb,Giudice:2004tc}.

Our discussion is based on supersymmetric theories in higher
dimensions. Crucial ingredients are GUT symmetry breaking by 
Wilson lines \cite{Witten:1985xc}, the generation of a fermion multiplicity by
magnetic flux \cite{Witten:1984dg} and the associated breaking of supersymmetry
\cite{Bachas:1995ik}. Interesting orbifold GUT models have been
constructed in five dimensions for $SU(5)$
\cite{Kawamura:2000ev,Hall:2001pg,Hebecker:2001wq} 
and in six dimensions for $SO(10)$ \cite{Asaka:2001eh,Hall:2001xr}. We consider supergravity in six dimensions
\cite{Nishino:1984gk, Nishino:1986dc} compactified on the orbifold $T^2/\mathbb{Z}_2$. Effects of
flux and Wilson lines,  in
particular the cancellation of anomalies due to the generated zero
modes, have recently been studied in \cite{Buchmuller:2015eya}. Magnetic flux also plays an important role in the stabilization of
the compact dimensions \cite{Braun:2006se}.

The proposed model is based on the gauge group $SO(10)\times U(1)_A$.
The three quark-lepton generations arise as zero modes of a bulk
$\bf16$-plet due to magnetic flux of the anomalous $U(1)_A$.
As a consequence, supersymmetry breaking is large, and squarks and
leptons are heavy. Following \cite{Asaka:2001eh,Hall:2001xr}, $SO(10)$ 
\cite{Georgi:1974my,Fritzsch:1974nn} is unbroken at one orbifold fixed point and broken at the other three 
to standard $SU(5)\times U(1)_X$ \cite{Georgi:1974sy},
the Pati-Salam group $SU(4)\times SU(2)\times SU(2)$ \cite{Pati:1974yy} and flipped
$SU(5)'\times U(1)_{X'}$ \cite{Barr:1981qv,Derendinger:1983aj}, respectively. The intersection of these
groups is the Standard Model gauge group, and the zero modes of bulk
fields uncharged under the anomalous $U(1)_A$ form 
$\mathcal{N} = 1$ gauge and Higgs split multiplets. Hence, at tree
level $\mathcal{N} = 1$ supersymmetry is unbroken in the gauge 
and Higgs sector.
 
In Sec.~\ref{sec:6DGUTs} the symmetry breaking of the $SO(10)$ GUT model will be
briefly reviewed. The effective supergravity actions in six and four
dimensions are discussed in Sec.~\ref{sec:GSMechanism}, following \cite{Buchmuller:2015eya}, with emphasis on
the cancellation of the $SO(10)\times U(1)_A$ anomaly induced by the
flux. Some aspects of the flavor structure of quark and lepton mass
matrices and quantum corrections to the mass spectrum are the topic
of Sec.~\ref{sec:Pheno}.

\section{\texorpdfstring{$\boldsymbol{SO(10)}$}{SO(10)} GUT in six dimensions}
\label{sec:6DGUTs}

Our starting point is a supersymmetric $SO(10)$ model in six
dimensions compactified on the orbifold $T^2/\mathbb{Z}_2$. In
addition to a vector multiplet in the adjoint representation $\bf 45$,
the model contains several hypermultiplets in the representations
$\bf 10$, $\bf 16$ and $\bf 16^*$.

A strong constraint on the consistency of the model is the cancellation
of bulk anomalies. The anomaly polynomial\footnote{In this paper we ignore gravitational anomalies.} is given by \cite{Erler:1993zy,Park:2011wv}
\begin{align}\label{ann1}
I^b_8 = \frac{\beta}{24} \left((2 - s_{\bf 10} + s_{\bf 16} + s_{\bf
    16^*})\text{tr} (\tilde{F}^4) + \frac{3}{16}(6-s_{\bf 10}) \text{tr} (\tilde{F}^2)^2\right) \,,
\end{align}
where $\beta= -1/(2\pi)^3$, $\tilde F$ is the $SO(10)$ field strength, and $s_{\bf 10}$,  $s_{\bf 16}$ and $s_{\bf
  16^*}$ are the multiplicities of the indicated representations. Note
that we have expressed all traces in terms of the trace in the $\bf
16$ representation, i.e.\ $\text{tr} \equiv \text{tr}_{\bf
  16}$. The components of the vector multiplet can be split into the
components of a 4d $\mathcal{N}=1$ vector multiplet
$A=(A_\mu,\lambda)$, $\mu = 0,\ldots, 3$,
and a chiral multiplet $\Sigma = (A_{5,6},\lambda')$, where $(\lambda,\lambda')$
forms a 6d Weyl fermion. Correspondingly, a hypermultiplet
$\phi$ splits into two chiral multiplets, $\phi = (\phi,\chi)$ and
$\phi^c = (\phi^c,\chi^c)$. Note that $\chi$ and $\chi^c$ are different, left-handed 4d Weyl
fermions. With respect to a $U(1)$ charge the hypermultiplet, one of
the chiral multiplets and the associated complex scalar carry the same
charge, hence these fields are all denoted by $\phi$. The second
chiral multiplet and the associated complex scalar carry opposite
charge and are therefore denoted by $\phi^c$. The orbifold compactification
breaks the $\mathcal{N} =2$ symmetry of the bulk to $\mathcal{N} =1$ via the boundary conditions
\begin{equation}\label{bc1}
\begin{alignedat}{2}
A(x,-y) &= A(x,y)\,, \quad & \Sigma (x,-y) &= - \Sigma (x,y) \,,\\
\phi_{\alpha}(x,-y) &= \eta^{\alpha} \phi_{\alpha}(x,y)\,, \quad &\phi_{\alpha}^c (x,-y) &= - \eta^{\alpha}
\phi_{\alpha}^c (x,y) \,,
\end{alignedat}
\end{equation}
where the $\eta^{\alpha}$ are parities of the fields $\phi_\alpha$ with $(\eta^{\alpha})^2=1$.
This breaking of supersymmetry at the fixed points generates
well-known fixed point anomalies. For $SO(10)$, however, they vanish since
$\text{tr} (\tilde F^3) = 0$.

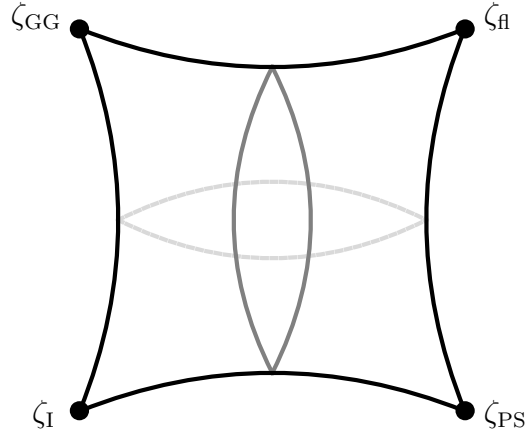
\begin{figure}[t]
\centering
\begin{overpic}[scale = 0.35, tics=10]{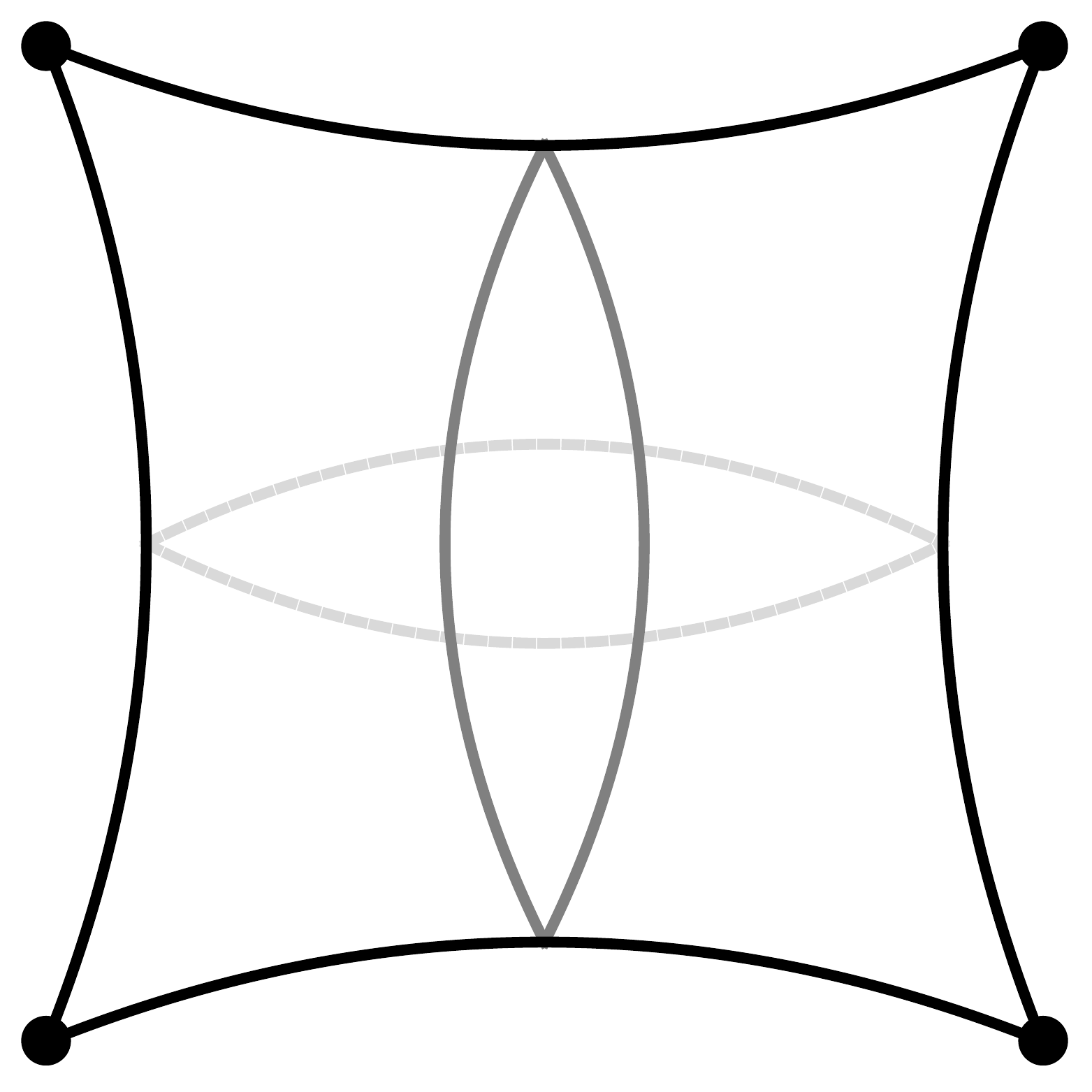}
	\put(-7, 2){$\zeta_{\mathrm{I}}$}
	\put(100, 2){$\zeta_{\mathrm{PS}}$}
	\put(100, 97){$\zeta_{\mathrm{fl}}$}
	\put(-12, 97){$\zeta_{\mathrm{GG}}$}
\end{overpic}
\caption{Structure of the gauge groups at the four fixed points. By choosing different boundary conditions, one obtains an unbroken $SO(10)$, a Georgi-Glashow $SU(5)$, a Pati-Salam, and a flipped $SU(5)$ GUT group at the four orbifold fixed points, respectively.}
\label{fig:GGStructure}
\end{figure}

A look at the anomaly polynomial \eqref{ann1} shows that a particular
choice of $SO(10)$ bulk fields is singled out: $s_{\bf 10} = 6$ and 
$s_{\bf 10} - s_{\bf 16} - s_{\bf 16^*} = 2$. In this case the entire
bulk anomaly vanishes. Such a GUT model has indeed  been studied,
with six $\bf10$-plets $H_1,\dots,H_6$, two $\bf16$-plets $\psi$, $\Psi$
and two $\bf16^*$-plets $\psi^c$, $\Psi^c$ \cite{Asaka:2002nd, Asaka:2003iy}. The breaking of the GUT group
$SO(10)$ takes place at the orbifold fixed points. $SO(10)$ remains
unbroken at the fixed point $\zeta_{\text{I}}=0$ whereas at the other three fixed points,
$\zeta_\text{PS}$, $\zeta_\text{GG}$ and $\zeta_\text{fl}$, $SO(10)$ is broken to standard
$SU(5)\times U(1)_X$, the Pati-Salam group $SU(4)\times
SU(2)\times SU(2)$ and flipped $SU(5)'\times U(1)_{X'}$, respectively (see Fig.~\ref{fig:GGStructure}). 
This is achieved by generalizing the boundary conditions \eqref{bc1}
\begin{equation}\label{bc2}
\begin{split}
P_iA(x,\zeta_i-y)P^{-1}_i   = \eta_iA(x,\zeta_i+y)\,, \quad &P_i\Sigma (x,\zeta_i-y)P^{-1}_i = - \eta_i\Sigma (x,\zeta_i+y) \,,\\
P_i\phi_\alpha(x,\zeta_i-y) = \eta^\alpha_i \phi_\alpha(x,\zeta_i+y)\,, \quad &P_i\phi_\alpha^c (x,\zeta_i-y) = - \eta^\alpha_i
\phi_\alpha^c (x,\zeta_i+y) \,,
\end{split}
\end{equation}
where $P_i$, $i \in \{\text{I,PS,GG,fl}\}$ are matrices breaking
$SO(10)$ to the respective subgroups.
The decompositions of the $SO(10)$ representations with respect to these 
subgroups read
\begin{align}
\begin{split}
\label{ps}
G_\text{PS}:\quad  &\bf45 \rightarrow (\bf15,\bf1,\bf1) \oplus (\bf1,\bf3,\bf1) \oplus (\bf1,\bf1,\bf3) \oplus (\bf6,\bf2,\bf2) \\
		   &\bf10 \rightarrow (\bf1,\bf2,\bf2) \oplus (\bf6,\bf1,\bf1)\\
		   &\bf16 \rightarrow (\bf4,\bf2,\bf1) \oplus (\bf4^*,\bf1,\bf2)\,, \quad \bf16^* \rightarrow (\bf4^*,\bf2,\bf1) \oplus (\bf4,\bf1,\bf2) 
\end{split}\\
\begin{split}
\label{gg}
G_\text{GG}, G_\text{fl}:\quad  &\bf45 \rightarrow \bf24_0 \oplus \bf1_0 \oplus \bf10_4 \oplus \bf10^*_{-4}\\
				&\bf10 \rightarrow \bf5_2 \oplus \bf5^*_{-2}\\
				&\bf16 \rightarrow \bf5^*_3 \oplus \bf10_{-1} \oplus \bf1_{-5}\,, \quad \bf16^* \rightarrow \bf5_{-3} \oplus \bf10^*_{1} \oplus \bf1_{5} 
\end{split}
\end{align}
The intersection of the groups $G_\text{PS}$, $G_\text{GG}$ and $G_{\text{fl}}$
contains the Standard Model group with an additional $U(1)$ factor,
$G_{\text{SM}}' = SU(3)\times SU(2) \times U(1)_Y\times U(1)_X$, and the
various hypermultiplet intersections yield Standard Model
representations. The parities $\eta^\alpha_i$ can be chosen such that
the zero modes of the $\bf10$-plets are Higgs doublets and 
color triplets, $H_1 \supset H_u$, $H_2 \supset H_d$, $H_{3,4} \supset D_{1,2}$,
$H_{5,6} \supset D^c_{1,2}$; the zero modes of one of the $\bf16$-plets
 and the $\bf16^*$-plets are weak doublets, color triplets and
singlets,$\psi \supset L$, $\psi^c \supset L^c$, $\Psi
\supset D^c, N^c$ and $\Psi^c \supset D, N$ \cite{Asaka:2002nd, Asaka:2003iy}. At each fixed point one
projects to a vector-like representation such that no fixed point
anomalies are generated. $H_{1,2}$ and $N^c\subset\Psi$, $N\subset\Psi^c$ play the role
of Higgs fields which break the electroweak symmetry and $B-L$,
respectively. The various vector-like exotics can become massive.

In the model described in \cite{Asaka:2002nd, Asaka:2003iy}, three $\bf16$-plets are
introduced at the 
three fixed points $\zeta_\text{PS}$, $\zeta_\text{GG}$
and $\zeta_\text{fl}$. They contain the quarks and leptons of the Standard
Model, in standard notation $\psi_i \rightarrow (q_i,l_i,
u^c_i,e^c_i,d^c_i,n^c_i)$. Hence, the chiral matter of the Standard Model is introduced as brane fields, 
unrelated to the bulk fields. 
In contrast, we shall pursue in the following a different approach in which the $\psi_i$ are not independent fields but rather
zero modes of the bulk field $\psi$ generated by the flux of an
anomalous $U(1)$. We therefore extend the bulk gauge group to
$SO(10)\times U(1)_A$, assign charge $q$ to $\psi$ and charge zero to all other
fields. The total gauge field is
\begin{align}
A = \tilde A^a T^a +  A' I\,, \quad F = dA + i A\wedge A = \tilde F + F'
\end{align}
and the covariant derivative for the $\bf16$-plet reads
$D_M = \partial_M + i\tilde A_M + iq A'_M$. This leads to a mixed bulk
anomaly. From the general expressions \cite{Erler:1993zy,Park:2011wv} one easily obtains
\begin{align}\label{ann2}
I^b_8 = \frac{\beta q^2}{24} \, \text{tr} \left(6 \, \tilde{F}\wedge \tilde F +  q^2 \, F'\wedge F'\right) \wedge
F'^2 \,.
\end{align}
In addition, a fixed point anomaly is generated,
\begin{align}\label{ann3}
I^f_8 = \frac{\alpha q}{24} \, \delta_O \, \text{tr}\left(3 \, \tilde F\wedge \tilde F  + q^2 \,
  F'\wedge F'\right) \wedge F'\wedge v_2 \,,
\end{align}
where $\alpha = 1/(2\pi)^2$, $\delta_O$ is a sum of $\delta$-functions
located at the four orbifold fixed points, and $v_2$ is the volume-form of the orbifold. The integrated anomaly polynomial
\begin{align}\label{ann4}
I_6 = \int_{T_2/\mathbb{Z}_2} I^f_8 = \frac{\alpha q}{24} \, 
\text{tr}\left(3 \, \tilde{F}\wedge\tilde F + q^2 \, F'\wedge F'\right)\wedge F'
\end{align}
corresponds to the 4d anomaly of a Weyl fermion that is a $\bf16$-plet
of $SO(10)$ with $U(1)_A$ charge $q$.

The bulk and fixed point anomalies \eqref{ann2} and \eqref{ann3} can be canceled by a generalization of the Green-Schwarz mechanism \cite{Green:1984sg}. The bulk part is factorizable and hence can be canceled in the standard way. Moreover, additional localized terms allow to cancel the fixed point anomalies, c.f.\ \cite{Scrucca:2004jn}.

Accounting for the $SO(10)$ symmetry breaking at $\zeta_i$, the fixed point
anomaly becomes (cf.~\cite{Asaka:2002my})
\begin{align}\label{ann6}
I^f_8 \propto \sum_{i=\text{I},\text{PS},\text{GG},\text{fl}} \delta_i \, \text{tr}\left(3 P_i \,
\tilde{F}\wedge\tilde F + q^2 \, F'\wedge F' \right) \wedge F'\wedge v_2 \,.
\end{align}
In order to cancel the additional contributions, further localized terms transforming in the various $SO(10)$ subgroups have to be included in the Green-Schwarz counter term.
We expect that it is possible 
to cancel all the gauge anomalies in this way. In fact,
there are examples of similar, anomaly free 6d supergravity theories, e.g.\ a 6d $SU(6)$ model that
was obtained as an intermediate step in a compactification of the
heterotic string \cite{Buchmuller:2007qf}. Note that also torus compactifications of Type I string theory can lead to the pattern of ``split supersymmetry", see \cite{Antoniadis:2004dt}. 
Since the focus of this paper is
on the additional zero modes generated by bulk magnetic flux, we shall ignore the effects of $SO(10)$ symmetry breaking on the fixed point anomalies in the following. A complete discussion will be given in \cite{bdrs1}.
 
\section{Flux and Green-Schwarz mechanism}
\label{sec:GSMechanism}

Let us now consider supergravity in six dimensions, following the discussion
in \cite{Buchmuller:2015eya}. The bosonic part of the 6d supergravity action
with gauge groups $SO(10)$ and $U(1)_A$ is given by
\begin{align}\label{sugra}
S = \int \left(\frac{1}{2} R - \frac{1}{2}d \phi \wedge \ast d \phi -
  \frac{1}{2}e^{2\phi} H \wedge \ast H 
- \frac{1}{2}e^{\phi}\text{tr}( \tilde F \wedge \ast \tilde F) - \frac{1}{2}e^{\phi} F' \wedge \ast F' \right) \,.
\end{align}
It involves the Ricci scalar $R$, the dilaton $\phi$, the field
strengths of the $SO(10)$ and $U(1)_A$ gauge fields $A = (\tilde A_M +
A'_M)dx^M$, and the field strength of the antisymmetric tensor field $B = \frac{1}{2} B_{MN} dx^M \wedge dx^N$,
\begin{align}
H = dB + X_3 \,.
\end{align}
Here $X_3$ is a linear combination of the $U(1)$ Chern-Simons term and localized contributions at the fixed points,
\begin{align}
X_3 = A' \wedge F'\, + \rho \, A' \delta_{O} v_2 \,,
\end{align}
with $\rho = \tfrac{\alpha}{2 q \beta}$.
Invariance of the action \eqref{sugra} under gauge
transformations, $\delta A = {\delta\tilde A + \delta A'} = d\tilde \Lambda + i[\tilde A,\tilde \Lambda]
+ d \Lambda'$, requires that the tensor field transforms as
\begin{align}
\delta dB = -\delta X_3 = - d \left(\Lambda' F' + \rho \, \Lambda' \delta_{O} v_2\right) \,.
\end{align}
Introducing the $SO(10)$ Chern-Simons 3-form
\begin{align}
\tilde\omega_3 = 
\text{tr}\left(\tilde A\wedge d\tilde A + \frac{2i}{3} \tilde A\wedge \tilde A \wedge \tilde A\right)\,,
\end{align}
the anomaly polynomials \eqref{ann2} and \eqref{ann3} correspond, up to local counter terms and a normalization, to the anomaly
\begin{equation}\label{A6}
\begin{split}
\mathcal A_6 = \enspace  \bar{\beta} \,  &\left( \tilde \omega_3 + \gamma A' \wedge F' \right) \wedge F' \wedge d\Lambda'  \\ 
		   {}+ \bar \alpha  &\left(\tilde \omega_3 + 2 \gamma A' \wedge F' \right) \wedge \delta_O v_2 \wedge d\Lambda'\,,
\end{split}
\end{equation}
where we have introduced the parameters $\bar{\beta} = 6 q^2 \beta$, $\bar{\alpha} = 3 q \alpha$, and $\gamma = \tfrac{q^2}{6} d_r$, with $d_r$ the dimension of the representation charged under the anomalous $U(1)_A$, here $d_r = 16$.
Note that $\mathcal{A}_6$ does not depend on $\tilde\Lambda$. This
fixes the Green-Schwarz counter term as
\begin{align}\label{GS}
S_{\text{GS}} = - \int \bar{\beta} \left(\,  \tilde{\omega}_3 + \gamma \, A' \wedge F' + \rho \gamma \, A' \delta_{O} v_2 \right) \wedge dB \,,
\end{align}
We now introduce a background field with constant flux,
\begin{align}
A' = \langle A' \rangle + \hat{A} \,,\quad F' = \langle F' \rangle +
\hat{F} \equiv f v_2 +\hat F\,.
\end{align}
Neglecting the dependence of the gauge fields $\tilde A$ and $\hat A$ on the
coordinates of the compact dimensions, one obtains from Eq.~\eqref{A6}
the 4d anomaly
\begin{align}
\mathcal A_4 = \int_{T_2/\mathbb{Z}_2} \mathcal{A}_6 = \frac{\bar \beta}{2} \left( f + 2 \rho \right) \left( \tilde \omega_3 + 2 \gamma \hat A \wedge \hat F \right) \wedge d\hat\Lambda\,.
\label{4dann}
\end{align}
It contains the effect of the 4d zero modes generated by the flux and the boundary conditions.
In a consistent truncation, where off-diagonal terms are set to zero, we decompose the redefined tensor field $\tilde B = B - \langle A'
\rangle \wedge \hat{A}$ \cite{Braun:2006se} as 
\begin{align}\label{dBtilde}
d \tilde{B} = db \wedge v_2 + d \hat{B} \,,
\end{align}
where $b$ is a real scalar field.
The axion transforms as  \({\delta db = - (f + 2 \rho) d \hat \Lambda} \) under 4d gauge transformations. 

With this truncation the decomposition of the field strength $H$ reads
\begin{align}\label{Hb}
H = (db +  f \, \hat{A} + \rho \delta_{O} \, \hat{A}) \wedge v_2 + \hat H\,, \quad
\hat{H} = d\hat{B} + \hat{A} \wedge \hat{F} \,.
\end{align}
Consequently, there is a $\delta_{O}^2$ contribution from the kinetic term of the 3-from $H$, which has to be regularized. In the following we use a regularization that is compatible with anomaly cancellation. A full description depends on the UV completion resolving the orbifold singularities and is beyond the scope of this paper.
 
It is now straightforward to evaluate the gauge part of the action \eqref{sugra} and the Green-Schwarz term \eqref{GS} in the case of
background flux, i.e.\ for the gauge fields $\tilde A$ and $\langle A' \rangle + \hat A$,
following \cite{Buchmuller:2015eya}. Performing dimensional reduction, replacing radion and dilaton by the real scalar fields
$t$ and $s$,
\begin{align}
t = r^2 e^{-\phi}\,, \quad s = r^2 e^\phi\,,
\end{align}
and dualizing the antisymmetric tensor $\hat B$ to the real scalar $c$,
\begin{align}\label{Hc}
\ast\hat{H} = \frac{1}{s^2} \left(dc + \tfrac{1}{2} \bar \beta \left(f+ 2 \rho \right) \gamma \hat{A}\right)\,,
\end{align}
one finally arrives at
\begin{align}
S_\mathrm{G} + S_\mathrm{GS} & 
		\begin{aligned}[t] =
			\int_{M\times X} \Big(&-\frac{1}{2} e^{2\phi} H \wedge \ast H - \frac{1}{2} e^\phi \text{tr} (\tilde F \wedge \ast \tilde F)  - \frac{1}{2} e^\phi F' \wedge \ast F' \\
			& {} -\bar{\beta} \left(\,  \tilde{\omega}_3 + \gamma \, A' \wedge F' + \rho \gamma \, A' \delta_{O} v_2 \right) \wedge dB \Big)
		\end{aligned} \nonumber \\
	     &
		\begin{aligned} \simeq
			\int_M\Big(&-\frac{s}{2} \, \text{tr} (\tilde F \wedge \ast \tilde F) -\frac{s}{2}\, \hat{F}\wedge\ast\hat{F} - \frac{f^2}{2t^2 s}  \\
		     &{} -\frac{1}{2t^2} \left(db + (f + 2\rho) \, \hat{A}\right)\wedge\ast\left(db + (f + 2\rho) \,\hat{A}\right)\\
		     &{} -\frac{1}{2s^2} \left(dc + \tfrac{1}{2} \bar \beta \gamma \left(f+ 2 \rho \right) \hat{A}\right)\wedge \ast \left(dc + \tfrac{1}{2} \bar \beta \gamma \left(f+ 2 \rho \right) \hat{A}\right) \\
		     &{} -\frac{1}{2}\, \bar{\beta} \, \left( \tilde{\omega}_3 + \gamma \hat A \wedge \hat F \right) \wedge d b - \hat{A} \wedge \hat{F} \wedge dc  \,.
		\end{aligned}
	\raisetag{\normalbaselineskip}
\label{4daction}
\end{align}
Here only the zero modes of $\tilde A$ contribute, which are contained
in the unbroken group $G'_{\text{SM}}$. Eqs.~\eqref{Hb} and \eqref{Hc} imply
for the 4d gauge transformation of the axion fields $\delta db = -
(f + 2 \rho) d\hat\Lambda$ and $\delta dc = -\tfrac{1}{2} \bar \beta \gamma \left( f + 2\rho \right) \, d\hat\Lambda$, respectively. 
One easily verifies that the total 4d action is gauge invariant, i.e.\ $\delta (S_\mathrm{G} + S_\mathrm{GS}) = \int\mathcal{A}_4$. Hence the chiral anomaly induced by the $U(1)_A$ flux is indeed canceled by the Green-Schwarz term.
  
For a bulk flux $f=-4 \pi N/q$ one obtains $N$ left-handed $SO(10)$ 16-plets $\psi_i$ as zero-modes. Their chiral anomaly is canceled by the Green-Schwarz term. After performing the Wilson-line breaking of $SO(10)$ to the Standard Model gauge group, an additional doublet $L$ associated with $\psi$ (see Sec.~\ref{sec:6DGUTs}) remains as a zero modes. It is not immediately obvious why this happens and why the $N$ 16-plets induced by the flux are not projected by the Wilson-line breaking. An important consistency check is the anomaly cancellation discussed above, and a more detailed picture is obtained by considering the zero-mode wave functions. For a $U(1)$ bulk flux the effect has been worked out in \cite{Buchmuller:2015eya}. The orbifold projection of $T^2$ to $T^2/\mathbbm{Z}_2$ yields for each 6d Weyl fermion one chiral 4d fermion. Without flux most zero modes of the 16-plet are projected out except for the doublet $L$. With flux one obtains $1+N$ zero modes for each mode that survives the Wilson-line breaking, and $N$ zero modes for each mode that is projected out by the Wilson lines. Altogether one then obtains $N$ 16-plets and one doublet ($L$) as zero modes. Note that the fields contained in the 16-plets develop different wave function profiles corresponding to their transformation properties with respect to the Standard Model subgroups of $SO(10)$. A detailed description of the $SO(10)$ wave functions will be given in \cite{bdrs1}.

The action \eqref{4daction} contains two axions, $b$ and $c$. 
One linear combination gives mass to the vector boson $\hat{A}$,
whereas a second linear combination, $a$, plays the role of a massless
axion.
The vector boson mass and the specific form of the linear combinations depend on the details of the regularization, the vacuum expectation values of the moduli fields $s$ and $t$, and the number of flux quanta $N = -\tfrac{qf}{4\pi}$.

The massless combination $a$ couples to the massive $U(1)$ vector boson $\hat A$ and to the
massless gauge fields of the Standard Model, as qualitatively described by the action
\begin{align}\label{SAa}
S_a =
\int_M\Big(-\frac{s_0}{2} \, \text{tr} (F_{\text{SM}'} \wedge\ast F_{\text{SM}'}) 
-\frac{\kappa}{2} \, da\wedge\ast da + \lambda a\, \text{tr} (F_{\text{SM}'}
\wedge F_{\text{SM}'})\Big)\,.
\end{align}
Note that $a$ receives a mass through non-perturbative QCD effects. Again, the parameters $\lambda$ and $\kappa$ are sensitive to the short distance behavior of the compactification.  For a pure $U(1)$ theory these quantities have been calculated in \cite{Buchmuller:2015eya}.

\section{Phenomenology}
\label{sec:Pheno}

In this section we briefly comment on phenomenological aspects of the
proposed model. A definite prediction is the flavor structure of the
quark and lepton mass matrices. At the different fixed points the
various Higgs fields are projected to representation of the respective
$SO(10)$ subgroups, 
\begin{align}
 \mathbf{10}\rightarrow\left\{
 \begin{array}{ll}
  H_1 \supset H_{\bf5}    \supset H_u,~H_2 \supset H_{\bf5^*} \supset H_d 	&\text{at } \zeta_\text{GG}\,,\\
  H_1 \supset H_{\tilde{\bf5}^*}  \supset H_u,~H_2 \supset H_{\tilde{\bf5}}   \supset H_d 	&\text{at } \zeta_\text{fl}\,,\\
  H_{1,2} \supset ({\bf1},{\bf2},{\bf2}) \supset \Delta_{1,2} = (H_u,H_d)			&\text{at } \zeta_\text{PS}\,,
 \end{array}
 \right.
\end{align}
where we have also indicated the doublets $H_u$ and $H_d$ of the
MSSM, which are contained as zero modes. Furthermore, we denote here and in the following the representations of $G_\text{fl}$ with a tilde in order to distinguish them from the representations of $G_\text{GG}$. The three ${\bf16}_i$-plets, $i=1,2,3$, of zero modes have the decomposition
\begin{align}
{\bf 16}_i\rightarrow\left\{
 \begin{array}{ll}
  ({\bf 5}^*_i, {\bf 10}_i, n^c_i)			&\text{at } \zeta_\text{GG}\,,\\
  (\tilde{{\bf 5}}^*_i, \tilde{{\bf 10}}_i, e^c_i)	&\text{at } \zeta_\text{fl}\,,\\
  ({\bf 4}_i,{\bf 4}^*_i)				&\text{at } \zeta_\text{PS}\,,
 \end{array}
 \right.
\end{align}
where we have suppressed the $U(1)$ charges and the $SU(2)\times
SU(2)$ transformation properties which are given in Eqs.~\eqref{gg} and \eqref{ps}, respectively.
The field $\Psi$, which transforms also in the ${\bf16}$ of $SO(10)$, decomposes in the same way. For later reference we introduce the following notation for some components of \(\Psi^*\):
\begin{align}
 \Psi^*\supset\left\{
 \begin{array}{r@{\;}ll}
  {\bf1}&= N					&\text{at } \zeta_\text{GG}\,,\\
  \tilde{\bf10}^*&=\tilde T^* \supset N		&\text{at } \zeta_\text{fl}\,,\\
  {\bf4}&=F \;\supset N				&\text{at } \zeta_\text{PS}\,.
 \end{array}
 \right.
\end{align}

At the fixed points, the $\mathcal{N} =2$ supersymmetry of the bulk is
broken to $\mathcal{N}=1$ supersymmetry. Hence, superpotential terms
of the type ${\bf 16_{~}16_{~}} H_{1,2}$ and ${\bf 16_{~} 16_{~}} \Psi^*
\Psi^*$ are allowed. They carry charge $2q$ with respect to $U(1)_A$, which
can be compensated in the standard way by an exponential term
involving the two axions. In the following we suppress the axion dependence in the quark and lepton couplings.

The fixed point superpotential
is determined by the symmetry breaking at $\zeta_i$,
\begin{align}
\label{Wfp}
W_\text{FP} =\ &\delta_{\text{I}~\;} (h^{\text{I}}_u {\bf16}_{~}{\bf16}_{~} H_1 + h^{\text{I}}_{d} {\bf16}_{~}{\bf16}_{~}H_2 + h^{\text{I}}_{n} {\bf16}_{~}{\bf16}_{~}{\Psi}^*{\Psi}^*)\nonumber\\
	     + &\delta_\text{GG} (h^\text{GG}_{u} {\bf10}_{~} {\bf10}_{~} H_{\bf5} + h^\text{GG}_{d} {\bf5}^*_{~}{\bf10}_{~} H_{\bf5^*} + h^\text{GG}_{\nu} {\bf5}^*_{~}n^c_{~} H_{\bf5} + h^\text{GG}_{n} n^c_{~}n^c_{~} NN)\nonumber\\
	     + &\delta_\text{PS} (h^\text{PS}_{u} {\bf4}_{~} {\bf4}^*_{~} \Delta_1 + h^\text{PS}_{d} {\bf4}_{~} {\bf4}^*_{~} \Delta_2 + h^\text{PS}_{n} {\bf4}^*_{~}{\bf4}^*_{~} F_{~} F_{~})\nonumber\\
	     + &\delta_\text{fl\;} (h^\text{fl}_{d} \tilde{\bf10}_{~} \tilde{\bf10}_{~} H_{\bf5} + h^\text{fl}_{u} \tilde{\bf5}^*_{~} \tilde{\bf10}_{~} H_{\bf5^*} + h^\text{fl}_{e}\tilde{\bf5}^*_{~} e^c_{~} H_{\bf5} + h^\text{fl}_{n} \tilde{\bf10}_{~}\tilde{\bf10}_{~} \tilde{T}^*\tilde{T}^*) \,.
\end{align}
At each fixed point, the superpotential couplings of the bulk fields
induce matrices $c_{ij}$ of couplings between the zero modes ${\bf16}_i$,
which are given by the products of the zero mode wave functions
at the respective fixed point. Since some wave functions vanish at
certain fixed points, the matrices $c_{ij}$ have a certain number of
zero entries. Hence, one obtains ``textures'' which are determined by
the local symmetry breaking patterns. From
Eq.~\eqref{Wfp} one obtains the 4d superpotential
\begin{align}
\label{eq:superpot}
W =\ & (h^{\text{I}}_u c^{\text{I}}_{ij} + h^\text{GG}_u c^\text{GG}_{ij} + h^\text{PS}_u c^\text{PS}_{ij} + h^\text{fl}_u c^\text{fl}_{ij})\;    H_u q_i u^c_j \nonumber\\
  +\ & (h^{\text{I}}_d c^{\text{I}}_{ij} + h^\text{GG}_d c^\text{GG}_{ij} + h^\text{PS}_d c^\text{PS}_{ij} + h^\text{fl}_d c^\text{fl}_{ij})\;    H_d q_i d^c_j \nonumber\\
  +\ & (h^{\text{I}}_d c^{\text{I}}_{ij} + h^\text{GG}_d c^\text{GG}_{ij} + h^\text{PS}_d c^\text{PS}_{ij} + h^\text{fl}_e c^\text{fl}_{ij})\;    H_d e^c_i l_j \nonumber\\
  +\ & (h^{\text{I}}_u c^{\text{I}}_{ij} + h^\text{GG}_\nu c^\text{GG}_{ij} + h^\text{PS}_u c^\text{PS}_{ij} + h^\text{fl}_d c^\text{fl}_{ij})\;  H_u l_i n^c_j \nonumber\\
  +\ & (h^{\text{I}}_n c^{\text{I}}_{ij} + h^\text{GG}_n c^\text{GG}_{ij} + h^\text{PS}_n c^\text{PS}_{ij} + h^\text{fl}_n c^\text{fl}_{ij})\;    n^c_i n^c_j N N \,.
\end{align}
Inserting the vacuum expectation values of the Higgs fields,
$\langle H_u \rangle = v_u$,  $\langle H_d \rangle = v_d$ and $\langle N \rangle = v_{B-L}$,
yields the quark and lepton mass matrices
\begin{align}
\mathcal{L}_m = \mathcal{M}^u_{ij} q_iu^c_j  + \mathcal{M}^d_{ij} q_id^c_j + \mathcal{M}^e_{ij} e^c_i l_j  +  \mathcal{M}^D_{ij} l_in^c_j  + \mathcal{M}^n_{ij} n^c_i n^c_j \,,
\end{align}
which can be read off from Eq.~\eqref{eq:superpot}. The detailed predictions for quark and lepton masses and the
CKM and PMNS mixing matrices will be described in \cite{bdrs1}.

For three quark-lepton generations the number of orbifold flux quanta
is $N=3$, and the masses of squarks and sleptons are given by \cite{Bachas:1995ik,Braun:2006se}
\begin{align}
M^2 = m^2_{\tilde q} = m^2_{\tilde l} = \frac{4\pi N}{V_2}\,,
\end{align}
where $V_2$ is the volume of the compact dimensions. Thus, $M^2$ is a
dynamical quantity which depends on the moduli fields. The corresponding moduli stabilization has to be
consistent with the unification of gauge couplings and proton decay.
At tree level,
gravitino, gauginos, higgsinos and Higgs bosons are massless.
The flux corresponds to a D-term breaking of supersymmetry \cite{Braun:2006se},
and quantum corrections will generate masses for all theses particles.
For $M \sim 10^{15}~\text{GeV}$, one would have $m_{3/2} \sim 10^{12}~\text{GeV}$.
Since at tree level gaugino masses are protected by an
R-symmetry and the higgsino masses by a PQ symmetry, one has 
\begin{align}
m_{\tilde q} = m_{\tilde l} \gg m_{3/2} \gg m_{1/2}, m_{\tilde h}\,.
\end{align}
This mass hierarchy is realized in split supersymmetry 
with gauginos in the TeV range \cite{ArkaniHamed:2004yi} or in
``spread supersymmetry'' with heavier gauginos and a higgsino LSP \cite{Hall:2011jd}.
The details of the mass spectrum depend on the treatment of quantum
corrections, in particular the contribution from anomaly mediation
\cite{Randall:1998uk,Giudice:1998xp}. Alternatively, one may be left just with the Standard Model and an axion.

A well-known problem of split supersymmetry is the fine-tuning of the
Higgs potential, not to mention the cosmological constant. It remains
to be seen whether the higher-dimensional framework discussed in this
paper can shed some new light on these problems.

\subsection*{Acknowledgments}

We thank Emilian Dudas, Koichi Hamaguchi and Satoshi Shirai for
helpful discussions.
This work has been supported by the German Science Foundation (DFG) within 
the Collaborative Research Center 676 ``Particles, Strings and the Early
Universe''. M.D. also acknowledges support from the Studienstiftung
des deutschen Volkes.

\appendix 
\setcounter{equation}{0}

\providecommand{\href}[2]{#2}\begingroup\raggedright\endgroup

\end{document}